\documentclass[12pt]{article}
\usepackage{amsmath,amssymb}
\usepackage{graphicx,color}
\numberwithin{equation}{section}
\usepackage{cite}
\usepackage{bm}
\usepackage{dcolumn}  
\newcommand{\beq}{\begin{equation}}
\newcommand{\eeq}{\end{equation}}
\newcommand{\beqa}{\begin{eqnarray}}
\newcommand{\eeqa}{\end{eqnarray}}
\newcommand{\beqar}{\begin{eqnarray*}}
\newcommand{\eeqar}{\end{eqnarray*}}
\newcommand{\al}{\alpha}

\renewcommand{\l}{\lambda}

\newcommand{\eg}{{\it e.g.,}\ }
\newcommand{\ie}{{\it i.e.,}\ }
\newcommand{\labell}[1]{\label{#1}} 
\newcommand{\reef}[1]{(\ref{#1})}
\newcommand\prt{\partial}

\begin{document}
\begin{titlepage}
\hfill
\vbox{
    \halign{#\hfil         \cr
           IPM/P-2007/038 \cr
           } 
      }  
\vspace*{20mm}
\begin{center}
{\Large {\bf  Entropy Function for Non-extremal \\
 D1D5 and D2D6NS5-branes}\\ }

\vspace*{15mm} 
\vspace*{1mm} 
{\large   Ahmad Ghodsi and Mohammad R. Garousi}
\vspace*{1cm}

{ {Department of Physics, Ferdowsi University, P.O. Box 1436, Mashhad, Iran}}\\
{ and}\\
{ Institute for Studies in Theoretical Physics and Mathematics (IPM)\\
P.O. Box 19395-5531, Tehran, Iran}\\

\vskip 0.6 cm

E-mail: {{ garousi@ipm.ir, ahmad@ipm.ir}} \\
\vspace*{1cm}
\end{center}

\begin{abstract} 
We apply the entropy function formalism to non-extremal $D1D5$ and $D2D6NS5$-branes 
whose throat approximation is given by the Schwarzschild black hole in 
$AdS_{3}\times S^{3}\times T^4$ and $AdS_{3}\times S^2\times S^{1}\times T^4$, respectively. 
We find the Bekenstein-Hawking entropy and the $(\alpha')^3R^4$ corrections 
from the value of the entropy function at its saddle point. While the higher derivative 
terms have no effect on the temperature, they decrease the value of the entropy.
\end{abstract} 

\end{titlepage} 

\section{Introduction} \label{intro} 
The black hole attractor mechanism has been an active subject over the past 
few years in string theory. This is originated  from the observation that there 
is a connection between the partition function of four-dimensional BPS black 
holes and partition function of topological strings \cite{Ooguri:2004zv}. 
This mechanism states that in the extremal black hole backgrounds 
the moduli scalar fields at horizon are determined by the charge of black 
hole and are independent of  their asymptotic values. 
One may study the attractor mechanism by finding the effective 
potential for the moduli fields and examining the behavior of the effective 
potential at its extremum. This extremum should be a local minimum for extremal 
black holes.  The entropy of black hole is 
then given by the value of the effective potential at its minimum. Using 
this, the entropy of some extremal black holes has been calculated in \cite{attractor}. 

Recently, it has been proposed  by A. Sen that 
the entropy of a specific class of extremal black holes in higher derivative 
gravity can be calculated using the entropy function formalism \cite{Sen}. 
According to this formalism, the entropy function for the black holes that 
their near horizon is $AdS_2\times S^{D-2}$ is defined by integrating the 
Lagrangian density over $S^{D-2}$ for a general $AdS_2\times S^{D-2}$ 
background characterized by the size of $AdS_2$ and $S^{D-2}$,  and taking 
the Legendre transform  of the resulting function with respect to the 
parameters labeling the electric fields. The result is  a function of   
moduli scalar fields as well as the sizes of $AdS_2$ and $S^{D-2}$. The 
values of moduli fields and the sizes are determined by extremizing 
the entropy function with respect to the moduli fields 
and the sizes. Moreover, the entropy is given by 
the value of the entropy function at the extremum\footnote{It is assumed 
that in the presence of higher derivative terms there is a solution whose near 
horizon geometry is $AdS_2\times S^{D-2}$. In the cases that the higher 
derivative corrections modify the solution such that the near horizon is 
not $AdS_2\times S^{D-2}$ anymore, one cannot use the entropy function formalism. 
In those cases one may use the Wald formula \cite{Wald} to calculated the 
entropy directly.}. Using this method the entropy of some extremal black 
holes have been found in \cite{Sen},\cite{Ghodsi:2006cd},\cite{entfunc}.

The above discussion does not indicate that the entropy function should have 
local minimum at the near horizon. In fact, it has been shown in \cite{mgag} 
that the entropy function has a saddle point at the near horizon of extremal 
black holes. One may then conclude that the entropy function formalism should 
not be something specific for the extremal black holes. Indeed, it has been 
shown in \cite{Cai:2007ik,mgag} that the entropy function formalism works for 
some non-extremal black hole/branes at the supergravity level. It has been 
speculated in \cite{mgag} that the entropy function  formalism works for the 
non-extremal black holes/branes whose near horizons are some extension of $AdS$ 
space, \eg  Schwarzschild black hole in $AdS$.

The non-extremal black branes that have been studied in \cite{mgag} 
are $D3$, $M2$ and $M5$-branes whose near horizon geometries are Schwarzschild 
black hole in $AdS_{p+2}$ where $p=3,2$ and $5$, respectively. When higher 
derivative corrections are included, however, the near horizon geometry is 
not the Schwarzschild black hole in $AdS_{p+2}$ anymore. Consequently, the 
entropy function formalism does not work for these cases when one considers 
the higher derivative terms. 
In this paper, we would like to study the non-extremal black hole/brane solutions 
that the higher derivative terms respect the symmetry of the tree level solutions. 
Consider the non-extremal $D1D5$ and $D2D6NS5$-branes. The near horizon (throat 
approximation) of their tree level geometries are the Schwarzschild black hole 
in $AdS_3$.  Moreover, in these cases, the higher derivative terms of the effective 
action respect the symmetry of the supergravity solution. In fact, the Schwarzschild 
black hole in $AdS_3$ is the BTZ black hole \cite{mb} in which the inner horizon 
$\rho_-=0$. On the other hand, it is known that the BTZ black hole is an exact 
solution of the string theory \cite{gh}. So one expects that the entropy function 
formalism works for the non-extremal $D1D5$ and $D2D6NS5$-branes even in the 
presence of the higher derivative terms.

An outline of the paper is as follows. In section 2, we review the non-extremal 
$D1D5$ and $D2D6NS5$ solutions of the effective action of type II string theory. 
In sections 3, using the entropy function formalism we derive the Bekenstein-Hawking  
entropy of $D1D5$-branes  in terms of the temperature of black branes. We  show that the entropy 
is given by the entropy function at its saddle point. In subsection 3.1  we show 
that the higher derivative terms respect the symmetries of the solution at the 
tree level and the entropy function formalism works in the presence of the 
higher derivative terms. Using this we find the entropy as the saddle point 
of the entropy function. As a double check, we also calculate the entropy 
using the Wald formula directly and find exact agreement with the result 
from the entropy function formalism. In section 4, we repeat the calculations 
for $D2D6NS5$-branes. We shall show that, in both cases, the higher derivative 
terms do not modify the tree level temperature, however, the entropy decreases 
with respect to the  Bekenstein-Hawking entropy.

\section{Review of the non-extremal solutions}
In this section we review the non-extremal $D1D5$ and $D2D6NS5$-branes solutions 
of the effective action of type II string theory. The two-derivatives effective 
action in the string frame is given by
\beqa 
S\!&=&\!\frac{1}{16\pi G_{10}}\int d^{10}x\,\sqrt{-g}\bigg\{
e^{-2\phi}\left(R+4(\prt\phi)^2-\frac{1}{12}H_{(3)}^2\right)
-\frac{1}{2}\sum\frac{1}{n!}F_{(n)}^2+\cdots\bigg\},
\eeqa 
where $\phi$ is the dilaton, $H_{(3)}$ is NS-NS 3-form field
strength, and $F_{(n)}$ is the electric R-R n-form field strength where 
$n=1,3,5$ for IIB and $n=2,4$ for type IIA theory.  In above equation, dots 
represent fermionic terms in which we are not interested. The effective action 
includes a Chern-Simons term which is zero for the $D1D5$ and $D2D6NS5$ solutions. Moreover, for these solutions $F_{(n)}=dC_{(n-1)}$. 
The 5-form field
strength tensor is self-dual, hence, it is not described by the
above simple action. It is sufficient to adopt the
above action for deriving the equations of motion, and impose the
self-duality by hand.

The non-extremal $D1D5$-branes solution of the IIB effective action when 
$D1$-branes are along the compact $(z)$ direction $(S^1)$ and $D5$-branes along the compact $(z,x_1,x_2,x_3,x_4)$ directions $(S^1\times T^4)$ is given 
by the following, (see e.g. \cite{D1D5}):
\beqa
ds^2&=&(f_1f_5)^{-\frac{1}{2}}\bigg(-fdt^2+dz^2\bigg)
+(f_1f_5)^{\frac12}\left(\frac{dr^2}{f}+r^2(d\Omega_{3})^2\right)
+\left(\frac{f_1}{f_5}\right)^{\frac12}\sum_{i=1}^4dx_i^2\,,\cr &&\cr
e^{-2\phi}&=&\frac{f_5}{f_1}\,\,,\qquad C_{tz}=\left(\frac{1}{f_1}-1\right)\,\,,\qquad
C_{tzx_1\cdots x_4}=\left(\frac{1}{f_5}-1\right)\,,
\eeqa
where
\beqa
f_1&=&1+\frac{Q_1}{r^2}\,,\qquad f_5=1+\frac{Q_5}{r^2}\,,\qquad f=1-\frac{r_0^2}{r^2}\,.
\eeqa
The above solution is the D1D5P solution \cite{D1D5,Maldacena} 	in which the  amount of left and right moving momenta, propagating in the compact direction $z$, is chosen to be equal, \ie  $\sigma=0$ in the notation \cite{Maldacena}. 

For $r_0=0$ we obtain the extremal solution, depending  on the two
parameters $Q_1$ and $Q_5$ which are  related to the number of $D$-branes. 
For $r_0\neq 0$ a horizon develops at $r=r_0$.  The near horizon geometry which 
is described by a throat, 
can be found by using the throat approximation where $r\ll Q_1$ and $r\ll Q_5$.
In these limits the non-extremal solution becomes 
\beqa
ds^2&=&\frac{r^2}{\sqrt{Q_1Q_5}}\bigg\{
-\bigg(1-\frac{r_0^2}{r^2}\bigg)dt^2
+dz^2\bigg\} +\frac{\sqrt{Q_1Q_5}}{r^2}
\left(1-\frac{r_0^2}{r^2}\right)^{-1}dr^2\cr &&\cr
&&+\sqrt{Q_1Q_5}(d\Omega_{3})^2+\sqrt{\frac{Q_1}{Q_5}}\sum_{i=1}^4dx_i^2,\cr &&\cr
e^{-2\phi}&=&\frac{Q_5}{Q_1}\,\,,\qquad F_{rtz}\,=\,2\frac{r}{Q_1}\,\,,
\qquad F_{rtzx_1\cdots x_4}\,=\,2\frac{r}{Q_5}\labell{Sch1}\,.
\eeqa 
The geometry is the product of $S^{3}\times T^4$ with the Schwarzschild 
black hole in $AdS_{3}$.
 
The  non-extremal $D2D6NS5$-branes solution of the IIA effective action 
when $D2$-branes are along the compact $(z,x_1)$ directions $(S^1\times S'^1)$, $D6$-branes along the compact $(z,x_1,x_2,x_3,x_4,x_5)$ directions $(S^1\times S'^1\times T^4)$ 
and $NS5$-branes along the compact $(z,x_2,x_3,x_4,x_5)$ directions $(S^1\times T^4)$  is given by the following 
(see e.g. \cite{ksks}):
\beqa
ds^2&=&(f_2f_6)^{-\frac{1}{2}}\bigg(-fdt^2+dz^2\bigg)
+f_5(f_2f_6)^{\frac12}\left(\frac{dr^2}{f}+r^2(d\Omega_{2})^2\right)\nonumber\\
&+&f_5(f_2f_6)^{-\frac12}dx_1^2+\left(\frac{f_2}{f_6}\right)^{\frac12}\sum_{i=2}^5
dx_i^2\,,\cr &&\cr
e^{-2\phi}&=&f_5^{-1}f_6^{\frac32}f_2^{-\frac12}\,\,,\qquad C_{tzx_1}=\coth\alpha_2
\left(\frac{1}{f_2}-1\right)+\tanh\alpha_2\,,\nonumber\\
H_{x_1ij}&=&\epsilon_{ijk}\prt_k f'_5\,\,,\qquad\,\,\,\,\, (dA)_{ij}
=\epsilon_{ijk}\prt_k f'_6\,,
\qquad i=6,7,8\,,
\eeqa
where
\beqa
f=1-\frac{r_0}{r}\,,\,\, f_n=1+\frac{r_0\sinh^2\alpha_n}{r}\,,\,\, f'_n=
1+\frac{r_0\sinh\alpha_n\cosh\alpha_n}{r}\,,\qquad n=2,5,6\,.
\eeqa
The above solution is the D2D6NS5P solution \cite{ksks,Maldacena} 	in which the  amount of left and right moving momenta, propagating in the compact direction $z$, is chosen to be equal, \ie  $\alpha_p=0$ in the notation \cite{Maldacena}.

For $r_0\rightarrow 0$ one obtains the extremal solution by sending 
$\alpha_n\rightarrow \infty$ such that $r_0\sinh^2\alpha_n\equiv Q_n$ 
is kept fixed. The extremal solution then  depends  on the three
 parameters $Q_2$, $Q_5$ and $Q_6$ which are  related to the number of 
 $D$-branes. 
For $r_0\neq 0$ a horizon develops at $r=r_0$.  The near horizon geometry 
which is described by a throat 
can be found by using the throat approximation where $r\ll Q_n$ and 
$Q_n\equiv r_0\sinh^2\alpha_n$.
In this limit $\cosh\alpha_n\sim\sinh\alpha_n$ and the non-extremal 
solution becomes 
\beqa
ds^2&=&\frac{\rho^2}{4Q_5\sqrt{Q_2Q_6}}\bigg\{
-\left(1-\frac{\rho_0^2}{\rho^2}\right)d\tau^2
+dy^2\bigg\} +\frac{4Q_5\sqrt{Q_2Q_6}}{\rho^2}
\left(1-\frac{\rho_0^2}{\rho^2}\right)^{-1}d\rho^2\cr &&\cr 
&+&Q_5\sqrt{Q_2Q_6}(d\Omega_{2})^2+\frac{Q_5}{\sqrt{Q_2Q_6}}dx_1^2+
\sqrt{\frac{Q_2}{Q_6}}\sum_{i=2}^5dx_i^2,\cr &&\cr
e^{-2\phi}&=&\frac{Q_6^{\frac32}}{Q_5\sqrt{Q_2}}\,,\,\, F_{\rho \tau yx_1}\,
=\,\frac{\rho}{2Q_5Q_2},\,\, H_{x_1\theta\phi}\,=\,-{Q_5}\sin\theta,\,\, 
(dA)_{\theta\phi}\,=\,-{Q_6}\sin\theta\,,\labell{Sch2}
\eeqa 
where we have made also the coordinate transformations 
$\tau=2\sqrt{Q_5}t,\,z=2\sqrt{Q_5}y,\,r=\rho^2$. The above geometry is now 
the product of $S^{2}\times S'^1\times T^4$ with the Schwarzschild black hole in $AdS_{3}$.


\section{Entropy function for non-extremal $D1D5$-branes}
Following \cite{Sen}, in order to apply  the entropy function formalism to the  non-extremal 
$D1D5$-branes one should  deform the near horizon geometry \reef{Sch1} to the most general form which is the product of the $AdS$-Schwarzchild and $S^3\times T^4$ space, that is
\beqa
ds^2_{10}&=&v_1\left[ \frac{r^2}{\sqrt{Q_1Q_5}}\bigg\{
-\left(1-\frac{r_0^2}{r^2}\right)dt^2
+dz^2\bigg\} +\frac{\sqrt{Q_1Q_5}}{r^2}
\left(1-\frac{r_0^2}{r^2}\right)^{-1}dr^2\right]\nonumber\\
&+&v_2 \left[ \sqrt{Q_1Q_5}(d\Omega_{3})^2+\sqrt{\frac{Q_1}{Q_5}}\sum_{i=1}^4
dx_i^2\right]\!,\cr &&\cr
e^{-2\phi}\!\!\!&=&\!\!\!\frac{Q_5}{Q_1}u\,\,, \qquad F_{rtz}\,=\frac{2r}{Q_1}
\frac{v_1^{\frac32}}{v_2^{\frac72}}\equiv\,e_1\,\,,\qquad F_{rtzx_1\cdots x_4}
\,=\frac{2r}{Q_5}v_1^{\frac32}v_2^{\frac12}\equiv\,e_2\,,\labell{deform}
\eeqa 
where $v_1,\,v_2,\,u$  are supposed to be constants, otherwise the above geometry is not product space.  The electric field strengths 
are deformed such that the electric charges are remaining fixed.  The function $f$ is 
defined to be the integral of the Lagrangian density over the 
horizon $H=S^1\times S^3\times T^4$. 
The result of inserting the background of \reef{deform} into $f$ is
\beqa 
f(v_1,v_2,u,e_1,e_2,r)&\equiv&\frac{1}{16\pi G_{10}}\int dx^H\sqrt{-g}\cal
L\cr &&\cr
&=&\frac{V_1V_3V_4r}{16\pi
G_{10}}Q_1^{3/2}Q_5^{-1/2}v_1^{3/2}v_2^{7/2}\nonumber\\
&\times &\left(\frac{6uQ_5^{\frac12}(v_1-v_2)}{Q_1^{\frac32}v_1v_2}+
\frac{Q_1^\frac12Q_5^\frac12}{2v_1^3r^2}e_1^2+\frac{Q_5^{\frac52}}
{2Q_1^{\frac32}v_1^3v_2^4r^2}e_2^2\right)\,,
\eeqa
where $V_1$ is the volume of $S^1$,  $V_3$ is the volume of  the 3-sphere with radius one, 
and $V_4$ is the $T^4$ volume. The electric charges are carried by the 
branes and are given by 
\beqa
q_1=\frac{\prt f}{\prt e_1}=\frac{V_1V_3V_4Q_1^2v_2^{\frac72}}
{16\pi G_{10}v_1^{\frac32}r}e_1\,,\qquad
q_2=\frac{\prt f}{\prt e_2}=\frac{V_1V_3V_4Q_5^2}
{16\pi G_{10}v_1^{\frac32}v_2^{\frac12}r}e_2\,.
\eeqa
Note that the electric charges are independent of the scales 
$v_1$ and $v_2$ as expected, \ie
\beqa
q_1=\frac{V_1V_3V_4}{8\pi G_{10}}Q_1\,, \qquad
q_2=\frac{V_1V_3V_4}{8\pi G_{10}}Q_5\,.\labell{charges}
\eeqa
Following \cite{Sen}, for $AdS_2$ space, one  defins the entropy function as the Legendre transform of $f$  
with respect to the electric fields $e_1$ and $e_2$. Extending that definition to our case which is $AdS_3$ space,  we define the entropy function by taking the Legendre transform of $f$  
with respect to the electric fields $e_1$ and $e_2$, and dividing the result by $r$, that is\footnote{For $AdS_{2+p}$ space, one should divide the Legendre transform of $f$ by $r^p$ to define the entropy function in $AdS_{2+p}$ space.}
\beqa 
F(v_1,v_2,u)&\equiv&\frac{1}{r}\left(e_1\frac{\prt f}{\prt e_1}+e_2\frac{\prt f}{\prt e_2}-f\right)\cr &&\cr
&=&\frac{V_1V_3V_4}{16\pi
G_{10}}v_1^{3/2}v_2^{7/2}\left(\frac{6u(v_2-v_1)}{v_1v_2}+\frac{2}{v_2^7}
+\frac{2}{v_2^3}\right)\,,
\eeqa
where we have substituted the values of $e_1$ and $e_2$ from  \reef{deform}. Note that we have already assumed that $v_1,\,v_2$ and $u$ are independent of $r$, that is, $\dot{v_1},\,\dot{v_2}$ and $\dot{u}$ are not appeared in the Lagrangian. Hence, diving the Legendre transform of $f$ by $r$ does not change the equations of motion\footnote{An alternative way to deal with the $AdS_{2+p}$ space is to dimensionally reduce it to $AdS_2$ space  and then use the entropy function formalism of the $AdS_2$ space \cite{Sahoo:2006vz}.}.  Solving the equations of motion 
\beqa
\frac{\prt F}{\prt v_i}\,=\,0\,, && i=1,2\,;\qquad \frac{\prt F}
{\prt u}\,=\,0\,,\labell{eom11}
\eeqa 
one finds the following solution
\beqa  
v_1\,=\,1\,,\,\, v_2\,=\,1\,,\,\, u=1.
\eeqa
This confirms that \reef{Sch1} is a solution of the type IIB supergravity action.

Let us now consider the behavior of the entropy function around the 
above critical point. To this end, consider the following matrix 
\beq
M_{ij}=\partial_{i}\partial_{j} F(v_1,v_2,u)\,.\labell{M1}
\eeq
Ignoring the overall  factor, the eigenvalues of this matrix are $(68.10,-10.87,0.78)$. 
This shows that the above critical point is a saddle point of the entropy function. 
It is a general property of the entropy function for both  extremal and 
non-extremal cases \cite{mgag}.

Let us now return to the entropy associated with this solution. 
It is straightforward to find the entropy from the Wald formula \cite{Wald}
\beqa 
S_{BH}&=&-\frac{8\pi}{16\pi G_{10}}\int dx^H\sqrt{
g^H}\frac{\prt {\cal L}}{\prt R_{trtr}}g_{tt}g_{rr}\,.\labell{wald10}
\eeqa 
For this background we have $R_{trtr}=\frac{1}{v_1\sqrt{Q_1Q_5}}g_{tt}g_{rr}$ 
and $\sqrt{-g}=v_1\sqrt{g^H}$.
These simplify the entropy relation to 
\beqa
S_{BH}=-\frac{8\pi \sqrt{Q_1Q_5}}{16\pi
G_{10}}\int dx^H\sqrt{-g}\frac{\prt
{\cal L}}{\prt R_{trtr}}R_{trtr}=-2\pi \sqrt{Q_1Q_5}\frac{\prt f_{\l}}{\prt
\l}\bigg|_{\l=1}\,,\labell{Waldlambda}
\eeqa 
where $f_{\l}$ is an expression similar to $f$
except that each $R_{trtr}$ Riemann tensor component is scaled by a factor of $\l$.

To find $\frac{\prt f_{\l}}{\prt\l}|_{\l=1}$ using the prescription given 
in \cite{Sen} and \cite{Ghodsi:2006cd}, 
we note that in addition to $R_{trtr}$, the  Riemann tensor components $R_{tztz}$ and
$R_{rzrz}$  are  proportional to $v_1$, \ie
\beqa
R_{trtr}=-\frac{v_1}{\sqrt{Q_1Q_5}}\,,\qquad
R_{rzrz}=\frac{v_1r^2}{\sqrt{Q_1Q_5}(r^2-r_0^2)}\,,\qquad
R_{tztz}=-\frac{v_1r^2(r^2-r_0^2)}{(Q_1Q_5)^{\frac32}}\,.\labell{4R}
\eeqa
Hence, one should rescale them too. We use the following scaling for these components  
\beq
R_{tztz}\rightarrow\l_1 R_{tztz}\,,\qquad R_{rzrz}\rightarrow\l_2 R_{rzrz}\,.\labell{3label} 
\eeq
Now, $f_{\l}(v_1,v_2,u,e_1,e_2)$ must be of the form $v_1^{\frac32}g(v_2,\l
v_1,\l_1v_1,\l_2v_1,e_1v_1^{-\frac32},e_2v_1^{-\frac32})$ for some function $g$.
Then one can show that the following relation holds for $f_\l$ and its derivatives 
with respect to scales, $\l_i,\, e_1,\, e_2$ and $v_1$:
\beqa
\l\frac{\prt f_{\l}}{\prt \l}+\l_1\frac{\prt
f_{\l}}{\prt \l_1}+\l_2\frac{\prt f_{\l}}{\prt
\l_2}+\frac{3}{2}e_1\frac{\prt
f_{\l}}{\prt e_1}+\frac{3}{2}e_2\frac{\prt
f_{\l}}{\prt e_2}+v_1\frac{\prt f_{\l}}{\prt
v_1}-\frac{3}{2}f_{\l}=0\,.\labell{xscald3}
\eeqa
In addition, there is a relation between the rescaled Riemann tensor 
components at the supergravity level, which can be found by using \reef{4R}
\beqa
\frac{\prt f_{\l}}{\prt \l_1}\bigg|_{\l_1=1}+\frac{\prt f_{\l}}{\prt
\l_2}\bigg|_{\l_2=1}=2\frac{\prt f_{\l}}{\prt \l}\bigg|_{\l=1}\labell{f1f2f}.
\eeqa
Replacing the above relation into \reef{xscald3} and using the equations of 
motion, one finds that $\frac{\prt
f_{\l}}{\prt \l}|_{\l=1}=-\frac{r}{2}F$. It is easy to see that the entropy 
is proportional to the entropy function up to a constant coefficient, \ie
\beqa 
S_{BH}=\pi\sqrt{Q_1Q_2}r_0F=\frac{V_1V_3V_4r_0\sqrt{Q_1Q_5}}{4G_{10}}\,,\labell{sbh}
\eeqa 
This  is the Bekenstein-Hawking entropy. One may write the entropy in terms 
of the temperature of black brane. The relation between $r_0$ and temperature can be read 
from the metric. The surface gravity is given by
\beq
\kappa=2\pi T=\sqrt{g^{rr}}\frac{d}{dr}\sqrt{-g_{tt}}\bigg|_H\labell{surg}
\eeq
which in our case we find $r_0=2\pi \sqrt{Q_1Q_5}T$. Note that the constant $v_1$ is canceled  in the above surface gravity. This causes that  the  higher derivative terms which modifies $v_1$  have no  effect on  the temperature. The entropy in terms of temperature becomes
\beqa
S_{BH}=2\pi N_1N_5V_1T\,,
\eeqa 
where we have used the relations $V_3=2\pi^2$, $V_4Q_1=16\pi^4\alpha'^3
g_sN_1$, $Q_5=\alpha' g_sN_5$, and
$16\pi G_{10}=(2\pi)^7\alpha'^4g_s^2$  where $N_1$ is the number of 
D1-branes and $N_5$ is the number of D5-branes \cite{Ghodsi:2006cd}. Alternatively, one may write the entropy in terms of the number of left moving or right moving momenta. Note that for our case $N_R=N_L$. The relation between $r_0$ and $N_R$  is given as
\beqa
N_R&=&\frac{r_0^2(V_1/2\pi)^2V_4/(2\pi)^4}{4g_s^2\alpha'^4}\nonumber\eeqa
where we have set $\sigma=0$ in the relations for $N_R$ and $N_L$ in \cite{Maldacena}. In terms of $N_R$, the entropy \reef{sbh} becomes\beqa
S_{BH}&=&4\pi\sqrt{N_1N_5N_R}\nonumber\\
&=&2\pi\sqrt{N_1N_5}\bigg(\sqrt{N_L}+\sqrt{N_R}\bigg)\eeqa
Note that for two charges  extremal black hole, $r_0=0$, \ie $N_R=N_L=0$, the entropy function is exactly 
the same as the non-extremal case but the value of the entropy is zero.

We have seen that the entropy function works  despite the fact that the 
horizon is not attractive. To see more explicitly  that the horizon here 
is not attractive, we use the intuitional explanation for attractor mechanism  
given in \cite{Kallosh:2006bt}. According to this, the physical distance 
from an arbitrary point to the attractive horizon is infinite. The proper 
distance of an arbitrary point from the horizon in our case is
\beq
\rho=\int_{r_0}^r \frac{(Q_1Q_5)^{1/4}}{r}(1-\frac{r_0^2}{r^2})^{-\frac12} 
dr=(Q_1Q_5)^{1/4} \log\bigg[\frac{r}{r_0}+\sqrt{\frac{r^2}{r_0^2}-1}\bigg]\,,
\eeq 
which is finite (infinite) for the non-extremal (extremal) case.

\subsection{Higher derivative terms }
In the previous sections we have seen that the entropy function formalism 
works at two derivatives level. It will be interesting to consider stringy 
effects and take a look at the entropy function mechanism again. To this 
end, we consider the higher derivative corrections coming from string theory. 
To next leading order the Lagrangian of type II theory is given by \cite{higher}
\beqa 
S=\frac{1}{16\pi G_{10}}\int d^{10}x\,\sqrt{-g}\bigg\{ {\cal L}^{tree}+
e^{-2\phi}\left(\gamma W\right)\bigg\}\,,\labell{eff1}
\eeqa 
where $\gamma=\frac18\zeta(3)(\al')^3$ and $W$ can be written in terms 
of the Weyl tensors
\beq
W=C^{hmnk}C_{pmnq}{C_h}^{rsp}{C^q}_{rsk}+\frac12 C^{hkmn}C_{pqmn}{C_h}^{rsp}{C^q}_{rsk}\,.
\eeq
Following \cite{Sen}, we consider the general background consist of $AdS$-Schwarzchild times $S^3\times T^4$ space \reef{deform} in the presence of the higher derivative terms. As we shall see shortly, the higher derivative terms respect the symmetry of the tree level solution, \ie the coefficients $v_1$ and $v_2$ remain constant. To see this we calculate the contribution of the above higher
derivative terms to the entropy function\footnote{Note that for $AdS_3\times S^3$ with identical radii, the Weyl tensor is zero as noted in \cite{Gubser:1998nz}. However, this tensor is non-vanishing for the space $AdS_3\times S^3\times T^4$ in which we are interested in 10-dimensional space-time.} 
\beqa 
\delta F\!\!&=&\!\!-\frac{\gamma Q_5u}{16\pi G_{10}rQ_1}\int dx^H\sqrt{-g}W= \cr &&\cr
\!\!&=&\!\!-\gamma u\frac{V_1V_3V_4\sqrt{Q_1Q_5}}{16\pi G_{10}}v_1^{\frac32}
v_2^{\frac72}\left[\frac{105(v_2^4-\frac47 v_1^3v_2+\frac{18}{35} v_1^2v_2^2
-\frac47 v_1v_2^3+v_1^4)}{32Q_1^2Q_5^2v_1^4v_2^4} \right]\!\!.
\eeqa
It is important to note that $\delta F$ is independent of $r$. This is consistent with our assumption that $v_1,\,v_2$ and $u$ are constants. By variation of  $F+\delta F$ with respect to $ v_1,\,v_2$ and $u$ one
finds the equations of motion. Since these equations are valid only up
to first order of $\gamma$, we consider the following perturbative
solutions:
\beqa 
v_1=1+\gamma x\,,\,\,\, v_2=1+\gamma y\,,\,\,\, u=1+\gamma z\,.\labell{v1v2}
\eeqa 
One should  replace them into the equations of motion, \ie  
\beqa
\frac{\prt(F+\delta F)}{\prt
u}=0\,\longrightarrow
&&6(y-x)=\frac{9}{2(Q_1Q_5)^{\frac32}}\,,\nonumber\\
\frac{\prt(F+\delta F)}{\prt v_1}=0\,\longrightarrow
&&28y+4x+8z=\frac{3}{(Q_1Q_5)^{\frac32}}\,,\cr &&\cr
\frac{\prt(F+\delta F)}{\prt v_2}=0\,\longrightarrow
&&-244y+84x-24z=-\frac{27}{(Q_1Q_5)^{\frac32}}  \,,
\eeqa
these equations are consistent and give the following results:
\beqa 
v_1=1-\gamma\frac{51}{32(Q_1Q_5)^{\frac32}}\,,\,\,v_2=1-\gamma\frac{27}
{32(Q_1Q_5)^{\frac32}}\,,\,\,
u=1+\gamma\frac{33}{8(Q_1Q_5)^{\frac32}}\,.\labell{sol1}
\eeqa
It is interesting to note that the stringy effect decreases the closed 
string coupling  at the near horizon, \ie $\phi=\phi_0-33\gamma/[16(Q_1Q_5)^{3/2}]$. 
Similar behavior appears for the non-extremal $D3$-branes \cite{Gubser:1998nz}. 

Let us now return to the entropy associated with this solution. The entropy is given by 
\beqa
S_{BH}=\pi\sqrt{Q_1Q_5}r_0(F+\delta F)\,,\labell{Waldlambdac}
\eeqa  
where we have used the fact that all the steps toward writing  the Wald 
formula \reef{Waldlambda} for  entropy in terms of the above entropy function 
remain unchanged. In particular the relation \reef{f1f2f} holds in the presence 
of the higher derivative terms. It turns out,  in order to find the entropy to linear order of $\gamma$, one 
does not need to know the values of $x,y$, and $z$. To see this,  note 
that if one replaces \reef{v1v2} into the first term above, one finds 
that $x,y$, and $z$ do not appear in this term linearly. The second term has 
an overall factor of $\gamma$, hence to the linear order of $\gamma$, one 
has to replace $v_1=v_2=u=1$ into it. The result is
\beqa
S_{BH}=\frac{V_1V_3V_4r_0\sqrt{Q_1Q_5}}{4G_{10}}\left[1-\gamma\frac{9}{8(Q_1Q_5)^{3/2}}
+O(\gamma^2)\right]\,.\labell{finalS}
\eeqa
As a double check, we calculate the entropy using the ward 
formula \reef{Waldlambda} directly, \ie
\beqa
S_{BH}=-2\pi\sqrt{Q_1Q_5}\left(\frac{\prt f_{\l}}{\prt\l}\bigg|_{\l=1}
+\frac{\prt f^W_{\l}}{\prt\l}\bigg|_{\l=1}\right)\,,
\eeqa
where the function $f^W$ is given by
\beq
f^W=\frac{\gamma}{16\pi G_{10}}\int dx^H \sqrt{-g} e^{-2 \phi}W\,.
\eeq
This second term is proportional to $\gamma$, so to the first order of 
$\gamma$ one has to replace the Schwarzschild $AdS$ solution \reef{Sch1} 
in ${\partial f_\l^W}/{\partial\l}$ which gives 
\beq
\frac{\partial f_\l^W}{\partial\l}\bigg|_{\l=1}=\gamma\frac{V_1V_3V_4r}
{16\pi G_{10}}\left[\frac{3}{(Q_1Q_5)^{3/2}}\right]\,.
\eeq
For the first term, on the other hand, one has to replace \reef{v1v2} which gives
\beq
\frac{\partial f_\l}{\partial\l}\bigg|_{\l=1}=\frac{V_1V_3V_4r}{16\pi G_{10}}
\left[-2-\gamma\frac{7y+x+2z}{(Q_1Q_5)^{3/2}}\right]\,.
\eeq
Now inserting the solution \reef{sol1} for $x,y$ and $z$ into the above equation, 
one finds exactly the result \reef{finalS}.

To write the entropy in terms of the temperature, we note that $v_1$ appears 
as an overall factor of $AdS_3$ in the background \reef{deform}, hence, the 
temperature \reef{surg} remains the same as the tree level temperature, \ie $r_0=2\pi\sqrt{Q_1Q_5}T$. This 
is unlike the temperature of non-extremal $D3$-branes that stringy effects 
increase the tree level temperature.

The entropy of $D1D5$-branes 
in terms of temperature or in terms of $N_R$ is
\beqa
S_{BH}&=&2\pi N_1N_5V_1T\left[1-\gamma\frac{9}{8}\left(\frac{(2\pi)^3V_4}{16\pi G_{10}N_1N_5}\right)^{3/2}+O(\gamma^2)\right]\nonumber\\
&=&4\pi\sqrt{N_1N_5N_R}\left[1-\gamma\frac{9}{8}\left(\frac{(2\pi)^3V_4}{16\pi G_{10}N_1N_5}\right)^{3/2}+O(\gamma^2)\right]\, .
\eeqa
In the second line, we have used the fact that the higher derivative corrections do not change the Hawking temperature which means the number of
excitations for the left and right moving momenta  remain the same as the tree level result. Note that the leading $\alpha'$ correction makes the entropy decreases. 


\section{Entropy function for non-extremal $D2D6NS5$-branes}
Following \cite{Sen}, in order to apply  the entropy function formalism to the  non-extremal 
$D2D6NS5$-branes one should  deform the near horizon geometry \reef{Sch2} to the most general form which is the product of the $AdS$-Schwarzchild and $S'^1\times S^2\times T^4$ space, that is  
\beqa
ds^2_{10}&=&v_1\left[ \frac{\rho^2}{4Q_5\sqrt{Q_2Q_6}}\bigg\{
-\left(1-\frac{\rho_0^2}{\rho^2}\right)d\tau^2
+dy^2\bigg\} +\frac{4Q_5\sqrt{Q_2Q_6}}{\rho^2}
\left(1-\frac{\rho_0^2}{\rho^2}\right)^{-1}d\rho^2 \right]\nonumber\\
&+&v_2 \left[ Q_5\sqrt{Q_2Q_6}(d\Omega_{2})^2+\frac{Q_5}{\sqrt{Q_2Q_6}}dx_1^2
+\sqrt{\frac{Q_2}{Q_6}}\sum_{i=2}^5dx_i^2 \right]\,,\cr &&\cr
F_{\rho \tau yx_1}&=&\frac{\rho}{2Q_5Q_2}\frac{v_1^{\frac32}}{v_2^{\frac52}}
\equiv\,e_1\,,\qquad H_{x_1\theta\phi}=-Q_5\sin\theta\,,\qquad (dA)_{\theta\phi}
=-Q_6\sin\theta\,,\cr &&\cr
e^{-2\phi}&=&\frac{Q_6^{\frac32}}{Q_5\sqrt{Q_2}}u\,,\labell{d2d6ns5}
\eeqa 
where $v_1,\,v_2,\,u$  are supposed to be constants.  The electric field strength 
is deformed such that the corresponding electric charge remains fixed. Similarly, 
to have the fixed magnetic charges, one does not need to deform the 
magnetic field strength. 
The function $f$ is defined to be the integral of the Lagrangian density over 
the horizon $H=S^1\times S'^1\times S^2\times T^4$. 
The result of inserting the background \reef{d2d6ns5} into $f$ is
\beqa 
f(v_1,v_2,,u,e_1)&\equiv&\frac{1}{16\pi G_{10}}\int dx^H\sqrt{-g}\cal
L\cr &&\cr
&=&\frac{V_1V'_1V_2V_4\rho}{32\pi
G_{10}}Q_2Q_5Q_6^{-1}v_1^{3/2}v_2^{7/2}\nonumber\\
&\times &\left(\frac{uQ_6(4v_1-3v_2)}{2Q_2Q_5^2v_1v_2}
+\frac{2Q_2Q_6}{v_1^3v_2\rho^2 }e_1^2-\frac{Q_6}{2v_2^2Q_5^2Q_2}
-\frac{Q_6u}{2v_2^3Q_5^2Q_2}\right)\,,
\eeqa
where $V_1(V'_1)$ is the volume of $S^1(S'^1)$,  $V_2$ is the volume of the 2-sphere with radius one, 
and $V_4$ is the $T^4$ volume. The electric charge carried by the $D2$-brane 
is given by 
\beqa
q_1=\frac{\prt f}{\prt e_1}=\frac{V_1V_1V_2V_4Q_2^2Q_5v_2^{\frac52}}
{8\pi G_{10}v_1^{\frac32}\rho}e_1\,.
\eeqa
Note that the electric charge is independent of the scales $v_1,v_2$ as expected, \ie
\beqa
q_1=\frac{V_1V_1V_2V_4}{16\pi G_{10}}Q_2\,.
\eeqa
Now we define the entropy function by taking the Legendre transform of $f$ 
with respect to the electric field $e_1$, and dividing by $\rho$, that is
\beqa 
F(v_1,v_2,u)&\equiv&\frac{1}{\rho}\left(e_1\frac{\prt f}{\prt e_1}-f\right)\cr &&\cr
&=&\frac{V_1V_1V_2V_4}{32\pi
G_{10}Q_5}v_1^{3/2}v_2^{7/2}\left(\frac{u(3v_2-4v_1)}{2v_1v_2}+\frac{1}
{2v_2^6}+\frac{1}{2v_2^2}+ \frac{u}{2v_2^3}\right)\,,\nonumber
\eeqa
where we have substituted the value of $e_1$. 
Solving the equations of motion
\beqa
\frac{\prt F}{\prt v_i}\,=\,0\,, && i=1,2\,;\qquad\frac{\prt F}{\prt u}\,
=\,0\labell{eom12}\,,
\eeqa 
one finds the following solutions
\beqa  
v_1\,=\,1\,,\,\, v_2\,=\,1\,,\,\, u=1\,.
\eeqa
This confirms that \reef{Sch2} is a solution of the type IIA supergravity action. 
To find the behavior of the entropy function around the above critical point, 
consider again the matrix \reef{M1}. 
Ignoring the overall factor, the eigenvalues of this matrix are $(12.44,-3.30,0.37)$. 
This shows again that the above critical point is a saddle point of the entropy function. 

Let us now return to the entropy associated with this solution. 
The Wald formula \cite{Wald} is given by
\beqa 
S_{BH}=-\frac{8\pi}{16\pi G_{10}}\int dx^H\sqrt{
g^H}\frac{\prt {\cal L}}{\prt R_{\tau\rho\tau\rho}}g_{\tau\tau}g_{\rho\rho}\,.
\labell{wald102}
\eeqa 
For this background we have $R_{\tau\rho\tau\rho}=\frac{1}{4v_1Q_5\sqrt{Q_2Q_6}}
g_{\tau\tau}g_{\rho\rho}$ and $\sqrt{-g}=v_1\sqrt{g^H}$.
These simplify the entropy relation to 
\beqa
S_{BH}=-\frac{32\pi Q_5\sqrt{Q_2Q_6}}{16\pi
G_{10}}\int dx^H\sqrt{-g}\frac{\prt
{\cal L}}{\prt R_{\tau\rho\tau\rho}}R_{\tau\rho\tau\rho}
=-8\pi Q_5\sqrt{Q_2Q_6}\frac{\prt f_{\l}}{\prt
\l}\bigg|_{\l=1}\,,\labell{Waldlambda2}
\eeqa 
where $f_{\l}$ is an expression similar to $f$
except that each $R_{\tau\rho\tau\rho}$ Riemann 
tensor component is scaled by a factor of $\l$.

To find $\frac{\prt f_{\l}}{\prt\l}|_{\l=1}$, 
we note that in addition to $R_{\tau\rho\tau\rho}$, the  Riemann 
tensor components $R_{\tau y\tau y}$ and
$R_{\rho y\rho y}$  are  proportional to $v_1$, \ie
\beqa
R_{\tau\rho\tau\rho}=-\frac{v_1}{4Q_5(Q_2Q_6)^\frac12}\,,\,
R_{\rho y\rho y}=\frac{v_1\rho^2}{4Q_5(Q_2Q_6)^\frac12(\rho^2-\rho_0^2)}\,,\,
R_{\tau y\tau y}=-\frac{v_1\rho^2(\rho^2-\rho_0^2)}{64Q_5^3(Q_2Q_6)
^{\frac32}}\,.\labell{4RR}
\eeqa
Hence, one should also rescale these components. We use the following scaling  
\beq
R_{\tau y\tau y}\rightarrow\l_1 R_{\tau y\tau y}\,,\qquad R_{\rho y\rho y}
\rightarrow\l_2 R_{\rho y\rho y}\,.\labell{3label2} 
\eeq
Now we see that $f_{\l}(v_1,v_2,u,e_1)$ must be of the form $v_1^{\frac32}g(v_2,\l
v_1,\l_1v_1,\l_2v_1,e_1v_1^{-\frac32})$ for some function $g$.
Then one can show that the following relation holds for $f_\l$ and its 
derivatives with respect to scales, $\l_i, e_1$ and $v_1$
\beqa
\l\frac{\prt f_{\l}}{\prt \l}+\l_1\frac{\prt
f_{\l}}{\prt \l_1}+\l_2\frac{\prt f_{\l}}{\prt
\l_2}+\frac{3}{2}e_1\frac{\prt
f_{\l}}{\prt e_1}+v_1\frac{\prt f_{\l}}{\prt
v_1}-\frac{3}{2}f_{\l}=0\,.\labell{xscald4}
\eeqa
As in the $D1D5$ case, by using the equation \reef{4RR} one finds the 
same relation as \reef{f1f2f} 
between the rescaled Riemann tensor components at the supergravity level. 
Replacing \reef{f1f2f}  into \reef{xscald4} and using the equations of 
motion, one finds again $\frac{\prt
f_{\l}}{\prt \l}|_{\l=1}=-\frac{\rho}{2}F$. Hence, the entropy is 
proportional to the entropy function up to a constant coefficient, \ie
\beqa 
S_{BH}=4\pi Q_5\sqrt{Q_2Q_6}\rho_0F=\frac{V_1V'_1V_2V_4\rho_0\sqrt{Q_2Q_6}}
{8G_{10}}\,,\labell{sbh2}
\eeqa 
This  is the Bekenstein-Hawking entropy. One may write the entropy 
in terms of  temperature. An alternative way to find temperature is to impose regularity of Euclidean metric near the horizon. So consider
the proper distance of an arbitrary point from the horizon, \ie
\beq
r=\int_{\rho_0}^{\rho} \frac{2(v_1Q_5)^\frac12(Q_2Q_6)^{\frac14}}{\rho}
(1-\frac{\rho_0^2}{\rho^2})^{-\frac12} d\rho=2(v_1Q_5)^\frac12(Q_2Q_6)^{\frac14} 
\log\bigg[\frac{\rho}{\rho_0}+\sqrt{\frac{\rho^2}{\rho_0^2}-1}\bigg]\,.
\eeq 
Near $\rho_0$, one finds $\rho^2=\rho_0^2(1+r^2/4v_1Q_5\sqrt{Q_2Q_6})$. So 
the metric \reef{d2d6ns5} near $\rho_0$ becomes 
\beqa
ds^2&=&-\frac{\rho_0^2}{16Q_5^2Q_2Q_6}r^2d\tau^2+dr^2+\cdots\,.
\eeqa
The period of the Euclidean time, required by the regularity of 
metric is  $1/T=\beta=8\pi Q_5\sqrt{Q_2Q_6}/\rho_0$. Note that here also the constant $v_1$ does not appear in the above metric, so the temperature is independent of the value of $v_1$. The 
entropy in terms of  the temperature is
\beqa
S_{BH}=\frac{\pi}{G_{10}}V_1V'_1V_2V_4Q_2Q_5Q_6T=2\pi N_2N_5N_6V_1T\,,
\eeqa 
where in the last expression we have used \beqa
N_2&=&\frac{1}{\sqrt{16\pi G_{10}}\mu_2}\int_{S^2\times T^4}\,*F_{(4)}\,
=\,\frac{Q_2V_2V_4}{16\pi G_{10}T_2}\,,\nonumber\\
N_5&=&\frac{g_s}{\sqrt{16\pi G_{10}}\mu_5}\int_{S^2\times S'^1}\,H_{(3)}\,
=\,\frac{g_sQ_5V'_1V_2}{16\pi G_{10}T_5}\,,\nonumber\\
N_6&=&\frac{1}{\sqrt{16\pi G_{10}}\mu_6}\int_{S^2}\,F_{(2)}\,
=\,\frac{Q_6V_2}{16\pi G_{10}T_6}\,,
\eeqa
where $\mu_p=\sqrt{16\pi G_{10}}T_p$ and $T_p=2\pi/((2\pi\ell_s)^{p+1}g_s)$. Alternatively, one may write the entropy in terms of the number of left moving or right moving momenta where in our case $N_R=N_L$. The relation between $\rho_0$ and $N_R$  is given as
\beqa
N_R&=&\frac{\rho_0^2(V_1/4\pi\sqrt{Q_5})^2V_4/(2\pi)^4V_1'/2\pi}{2g_s^2\alpha'^4}\nonumber\eeqa
where we have set $\alpha_p=0$ in the relations for $N_R$ and $N_L$ in \cite{Maldacena}, and used the rescaling $z=2\sqrt{Q_5}y$, $r_0=\rho_0^2$. In terms of $N_R$, the entropy \reef{sbh2} becomes\beqa
S_{BH}&=&4\pi\sqrt{N_2N_6N_5N_R}\nonumber\\
&=&2\pi\sqrt{N_2N_6N_5}\bigg(\sqrt{N_L}+\sqrt{N_R}\bigg)\eeqa which is in the conventional form appearing in \cite{Maldacena}.
\subsection{Higher derivative terms }
We now consider the general background consist of $AdS$-Schwarzchild times $S'^1\times S^2\times T^4$ space \reef{d2d6ns5} in the presence of the higher derivative terms. The higher derivative terms respect the symmetry of the tree level solution, \ie the coefficients $v_1$ and $v_2$ remain constant. To see this we calculate the contribution of the above higher
derivative terms to the entropy function \ie
\beqa 
\delta F&=&-\frac{\gamma Q_6^{\frac{3}{2}}u}{16\pi G_{10}Q_5Q_2^\frac12\rho}
\int dx^H\sqrt{-g}W=-\frac{\gamma u V_1V'_1V_2V_4(Q_2Q_6)^\frac12v_1^
{\frac32}v_2^{\frac72}}{32\pi G_{10}}\times\cr &&\cr
&\times&\left[\frac{35(-\frac{3}{28}v_1^3v_2+\frac{81}{2048}v_2^4+
\frac{27}{224}v_2^2v_1^2-\frac{27}{896}v_2^3v_1+v_1^4)}
{108v_1^4v_2^4(Q_2Q_6)^2Q_5^4} \right]\,,
\eeqa
By variation of $F+\delta F$ with respect to $v_1,\, v_2$ and $u$ one
finds the equations of motion. Considering the perturbative
solutions \reef{v1v2}, one finds 
\beqa
\frac{\prt(F+\delta F)}{\prt
u}=0\,\longrightarrow
&&y-3x=\frac{73315}{110592(Q_2Q_5^2Q_6)^{\frac32}}\,,\nonumber\\
\frac{\prt(F+\delta F)}{\prt v_1}=0\,\longrightarrow
&&7y+x+2z=-\frac{7075}{12288(Q_2Q_5^2Q_6)^{\frac32}}\,,\cr &&\cr
\frac{\prt(F+\delta F)}{\prt v_2}=0\,\longrightarrow
&&41y-21x+2z=-\frac{44395}{110592(Q_2Q_5^2Q_6)^{\frac32}}  \,,
\eeqa
these equations are consistent, and give the following results
\beqa 
v_1&=&1-\gamma\frac{247343}{884736(Q_2Q_5^2Q_6)^{\frac32}}\,,\quad
v_2=1-\gamma\frac{155509}{884736(Q_2Q_5^2Q_6)^{\frac32}}\,,\cr &&\cr
u&=&1+\gamma\frac{45917}{98304(Q_2Q_5^2Q_6)^{\frac32}}\,.
\eeqa
Similar to the $D1D5$ case, the stringy effects decrease the 
closed string coupling at the near horizon.
Let us return to the entropy associated with this solution. The entropy is given by 
\beqa
S_{BH}=4\pi Q_5\sqrt{Q_2Q_6}\rho_0(F+\delta F)\,, \labell{Waldlambdac2}
\eeqa  
where again we have used the fact that all the steps toward writing the Wald formula 
for the entropy in terms of the entropy function above, remain unchanged. 
In this case also, in order to find the entropy to linear order of $\gamma$, one does not need to know the solutions 
for $x,y$ and $z$.  That is, if one replaces \reef{v1v2} into the tree level entropy 
function, \ie the first term above, one finds that $x,y$, and $z$ do not appear in it linearly.   
The second term, on the other hand,  has an overall factor of $\gamma$, hence to 
linear order of $\gamma$, one has to replace $v_1=v_2=u=1$ into it. The result is
\beqa
S_{BH}=\frac{V_1V'_1V_2V_4\rho_0(Q_2Q_6)^\frac12}{8G_{10}}
\left[1-\gamma\frac{73315}{221184(Q_2Q_5^2Q_6)^{\frac32}}+O(\gamma^2)\right]\,,
\labell{finalS2}
\eeqa
As a double check, we calculate the entropy using the ward formula 
\reef{Waldlambda2} directly, \ie
\beqa
S_{BH}=-8\pi Q_5\sqrt{Q_2Q_6}\left(\frac{\prt f_{\l}}{\prt\l}\bigg|_{\l=1}
+\frac{\prt f^W_{\l}}{\prt\l}\bigg|_{\l=1}\right)\,.
\eeqa
The second term is proportional to $\gamma$, so to the first order 
of $\gamma$ one has to replace the Schwarzschild $AdS$ solution \reef{Sch2} 
in ${\partial f_\l^W}/{\partial\l}$ which gives 
\beq
\frac{\partial f_\l^W}{\partial\l}\bigg|_{\l=1}=\gamma\frac{V_1V'_1V_2V_4\rho}
{16\pi G_{10}Q_5}\left[\frac{1205}{110592(Q_2Q_5^2Q_6)^{\frac32}}\right]\,.
\eeq
For the first term, one has to replace \reef{v1v2} which gives
\beq
\frac{\partial f_\l}{\partial\l}\bigg|_{\l=1}=\frac{V_1V'_1V_2V_4\rho}{16\pi G_{10}Q_5}\left[-\frac{1}{4}-\gamma\frac{7y+x+2z}{8(Q_2Q_5^2Q_6)^{\frac32}}\right]\,.
\eeq
Now inserting the solutions for $x,y$ and $z$ into the above equation, 
one finds exactly the result \reef{finalS2}.
The entropy \reef{finalS2} in terms of  temperature is
\beqa
S_{BH}=2\pi N_2N_5N_6V_1 T\left[1-\gamma\frac{73315}{221184(Q_2Q_5^2Q_6)^{3/2}}
+O(\gamma^2)\right]\,.
\eeqa
This entropy, like the entropy of the $D1D5$-branes, is less than the 
Bekenstein-Hawking entropy. This is unlike the entropy of the non-extremal 
$D3$-branes \cite{Gubser:1998nz} which is $
S_{BH}^{D3}=\frac{\pi^2}{2}N^2V_3T^3(1+15\gamma+O(\gamma^2))$, where the 
first term is the Bekenstein-Hawking entropy and the second term which is the $\alpha'$ correction, is positive.

The increase in the entropy for $D3$-branes is consistent with the fact that the Bekenstein-Hawking entropy at strong 't Hooft coupling is less than the  entropy of $N=4$ SYM theory at zero coupling by a factor of $3/4$ \cite{Gubser:1996de}. On the other hand, the correction to the entropy at  weak coupling is negative \cite{Fotopoulos:1998es} which is an indication of smooth interpolation between the weak and strong coupling regimes. For $D1D5$-branes, our result indicates that the correction to the entropy at strong coupling is negative. On the other hand, it is known that the entropy at zero coupling is the same as the Bekenstein-Hawking entropy at strong coupling \cite{Callan:1996dv}. This indicates that the correction to the entropy at weak coupling should be non-vanishing too. Assuming the   interpolating function  between the strong and the weak coupling regimes of the Higgs branch of the $N=(4,4)$ SYM at finite temperature in $1+1$ dimensions does not cross the zeroth order entropy at any point in finite coupling, one expects  the correction to the entropy at weak coupling to be negative. It would be interesting to perform this calculation.


\end{document}